\documentclass[fleqn,usenatbib]{mnras}

\usepackage{newtxtext,newtxmath}

\usepackage[T1]{fontenc}

\DeclareRobustCommand{\VAN}[3]{#2}
\let\VANthebibliography\thebibliography
\def\thebibliography{\DeclareRobustCommand{\VAN}[3]{##3}\VANthebibliography}

\usepackage{graphicx}	
\usepackage{layouts}

\title[Nested sampling with any prior you like]{Nested sampling with any prior you like}

\author[J. Alsing and W. Handley]{
Justin Alsing$^{1,2}$\thanks{E-mail: justin.alsing@fysik.su.se} and
Will Handley$^{3,4}$
\\
$^{1}$Oskar Klein Centre for Cosmoparticle Physics, Department of Physics, Stockholm University, Stockholm SE-106 91, Sweden\\
$^{2}$Imperial Centre for Inference and Cosmology, Department of Physics, Imperial College London, Blackett Laboratory,Prince Consort Road, London SW7 2AZ, UK\\
$^{3}$Astrophysics Group, Cavendish Laboratory, JJ Thomson Avenue, Cambridge CB3 0HA, UK\\
$^{4}$Kavli Institute for Cosmology, Madingley Road, Cambridge CB3 0HA, UK
}
\date{Accepted XXX. Received YYY; in original form ZZZ}

\pubyear{2020}

\begin{document}
\label{firstpage}
\pagerange{\pageref{firstpage}--\pageref{lastpage}}
\maketitle

\begin{abstract}
Nested sampling is an important tool for conducting Bayesian analysis in Astronomy and other fields, both for sampling complicated posterior distributions for parameter inference, and for computing marginal likelihoods for model comparison. One technical obstacle to using nested sampling in practice is the requirement (for most common implementations) that prior distributions be provided in the form of transformations from the unit hyper-cube to the target prior density. For many applications -- particularly when using the posterior from one experiment as the prior for another -- such a transformation is not readily available. In this letter we show that parametric bijectors trained on samples from a desired prior density provide a general-purpose method for constructing transformations from the uniform base density to a target prior, enabling the practical use of nested sampling under arbitrary priors. We demonstrate the use of trained bijectors in conjunction with nested sampling on a number of examples from cosmology. 
\end{abstract}

\begin{keywords}
data analysis: methods
\end{keywords}


\section{Introduction}
Nested sampling has become one of the pillars of Bayesian computation in Astronomy (and other fields), enabling efficient sampling of complicated and high-dimensional posterior distributions for parameter inference, and estimation of marginal likelihoods for model comparison \citep{Skilling_2006,Feroz_2009,Handley_2015a, Handley_2015b}. In spite of its widespread popularity, nested sampling has had a practical limitation owing to the fact that in several implementations priors must typically be specified as a transform from the unit hyper-cube. As a consequence, nested sampling has largely only been practical for a somewhat restrictive class of priors, which have a readily available representation as a transform from the unit hyper-cube.

This limitation has been notably prohibitive for (commonly occurring) situations where one wants to use an interim posterior derived from one experiment as the prior for another. In these situations, while the interim posterior is computable up to a normalization constant so can be sampled from, no unit hyper-cube transform representation readily presents itself in general.

In this letter we address the need to provide priors in the form of transformations from the unit hyper-cube, by training parametric bijectors\footnote{Note that the required transformations do not always need to be bijective (ie., invertible), for example in the case of discrete parameters. We focus on continuous parameters in this paper.} to represent priors given only a set of samples from the desired prior. For use cases where interim posteriors are to be used as priors, the bijector can be trained on a set of samples from the interim posterior. In the context of cosmological data analysis, representing interim posteriors as (computationally inexpensive) bijectors has an additional advantage: it circumvents the need to either multiply computationally expensive likelihoods together, or introduce an extra importance sampling step, when performing joint analyses of multiple datasets. This will hence lead to substantial computational (and convenience) gains when performing multi-probe cosmological analyses.

We note that whilst most commonly used nested sampling implementations require priors as unit hyper-cube transforms (e.g., \texttt{multinest} \citep{Feroz_2009} and \texttt{polychord} \citep{Handley_2015a, Handley_2015b}), it is not a \emph{theoretical} requirement. Early implementations used gaussian random walk MCMC to draw prior samples, and some implementations still use alternative prior sampling strategies (eg., \texttt{dnest4} \citealp{brewer2016}). Regardless of implementation, if samples from one sampling run are to be used to construct a prior for a subsequent run, some additional engineering (such as the method presented in this letter) is required.

The structure of this letter is as follows. In \S \ref{sec:nested} we briefly review nested sampling theory. In \S \ref{sec:bijectors} we discuss representation of priors as bijective transformations from a base density. In \S \ref{sec:cosmological_bijectors} we use bijectors for representing cosmological parameter posteriors for a number of experiments, demonstrating that they are sufficiently accurate for practical use as priors for subsequent (multi-probe) cosmological analyses. We conclude in \S \ref{sec:conclusions}.
\section{Nested sampling and prior specification}
\label{sec:nested}
Given a likelihood function $\mathcal{L}(\theta)$ and prior distribution $\pi(\theta)$, nested sampling was incepted by \citet{Skilling_2006} as a Monte Carlo tool to compute the Bayesian evidence
\begin{equation}
    \mathcal{Z} = \int \mathcal{L}(\theta)\pi(\theta) d\theta,
    \label{eqn:evidence}
\end{equation}
whilst simultaneously generating samples drawn from the normalised posterior distribution $\mathcal{P}(\theta) = \mathcal{L}(\theta)\pi(\theta)/\mathcal{Z}$.

Nested sampling performs these dual processes by evolving an ensemble of $n_\mathrm{live}$ \emph{live points} by initially drawing them from the prior distribution $\pi$, and then at each subsequent iteration: (a) discarding the lowest likelihood live point and (b) replacing it by a live point drawn from the prior $\pi$ subject to hard constraint that the new point is at a higher likelihood than the discarded point. The set of discarded \emph{dead points} and likelihoods can be used to compute a statistical estimate of the Bayesian evidence \eqref{eqn:evidence}, and be weighted to form a set of samples from the posterior distribution.

The nested sampling procedure therefore differs substantially from traditional strategies such as Gibbs sampling, Metropolis-Hastings or Hamiltonian Monte Carlo in that it is acts as a likelihood scanner rather than a posterior sampler. It scans athermally, so in fact is capable of producing samples at any temperature and computing the full partition function in addition to the Bayesian evidence~\citep{Handley_2019}.

Implementations of the nested sampling meta-algorithm e.g.\ \texttt{MultiNest}~\citep{Feroz_2009} and \texttt{PolyChord}~\citep{Handley_2015a,Handley_2015b} differ primarily in their choice of method by which to draw a new live point from the prior subject to a hard likelihood constraint. Nested sampling algorithms typically assume that each parameter has a prior that is uniform over the prior range $[0,1]$ i.e. the \emph{unit hypercube}. As we shall see in the next section, this is not as restrictive as it sounds, since any proper prior may be transformed onto the unit hypercube via a bijective transformation.

It should be noted that normalising flows and bijectors have been used by \cite{2020MNRAS.496..328M} \& \cite{2021arXiv210211056W} in the context of techniques for generating new live points, but the approach in this letter for using them for prior specification is applicable to all existing nested sampling algorithms.

\begin{figure*}
    \centerline{\includegraphics{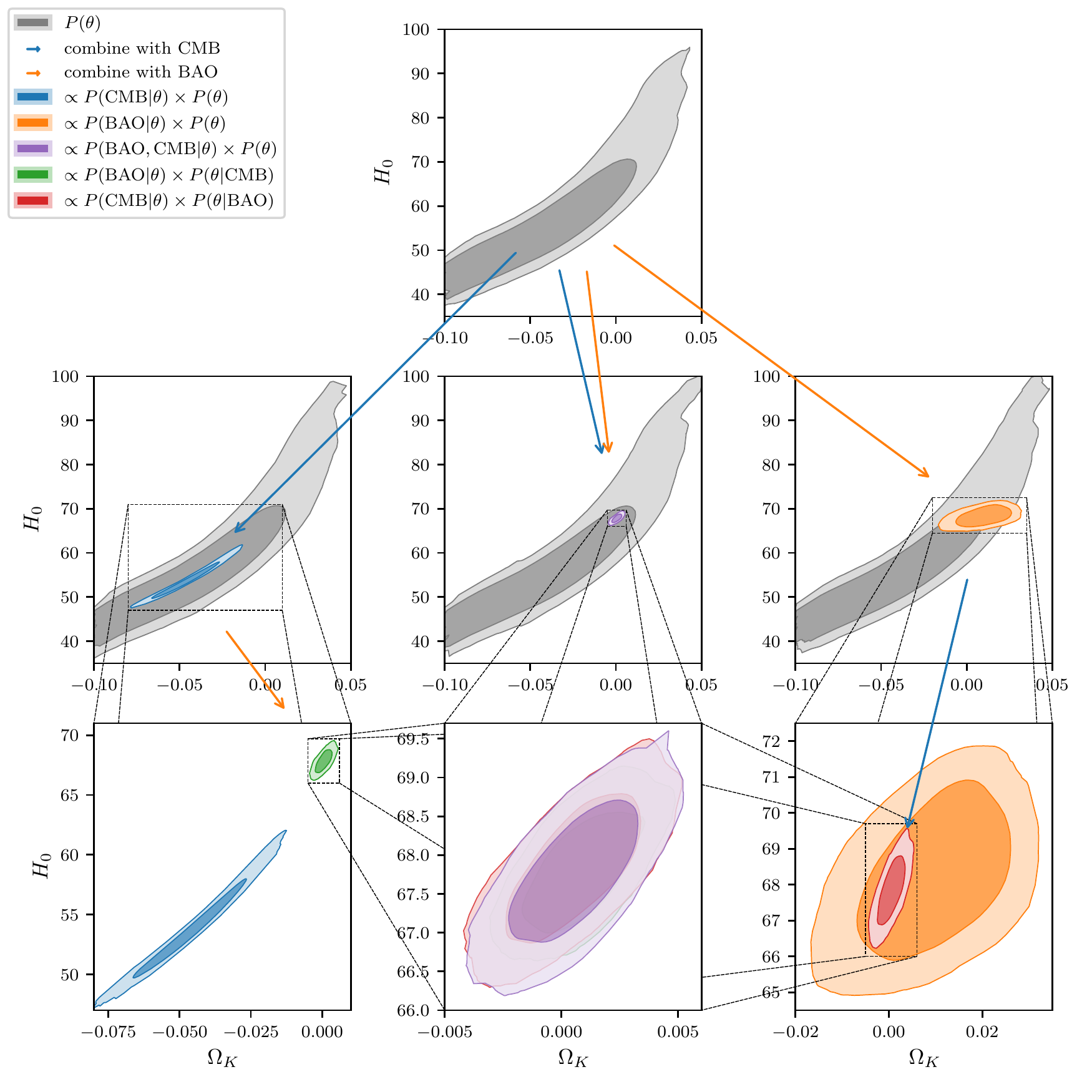}}
    \caption{Demonstration of consistency between sampling using traditional and bijector-based priors. All plots show $66.7\%$ and $95\%$ credibility regions of marginal probability distributions projected into the $\Omega_K$-$H_0$ plane, and colours are chosen to be consistent across the grid. The first row indicates the prior placed on the $\Omega_K$ $H_0$ parameter space by \texttt{CAMB} ruling certain portions of the parameter space to be unphysical (e.g. for Universes with too much negative curvature collapsing, or unphysical lensing amplitudes), and it should be noted that it is banana-like rather than flat. The second row indicates the prior from the first row, and the posterior distributions derived by using CMB data alone, CMB+BAO data and BAO data alone respectively for each column, with appropriately coloured arrows indicating the application of data to the prior to produce a corresponding posterior distribution. The final row demonstrates the comparison between approaches. In the first and final columns, we see the resultant posterior derived by using the corresponding posterior from the second row as a prior for the cosmology run. The central bottom panel compares the three routes to a combined data posterior and demonstrates them to be in general agreement to within sampling error. Dashed lines indicate the corresponding relationship between limits of the plots. Plot produced under \texttt{anesthetic} \citep{Handley_2019}.\label{fig:cosmology_posterior}}
    \vspace{50pt}
\end{figure*}

\section{Representing priors as bijectors}
\label{sec:bijectors}
A bijector $\mathbf{f}: \mathbf{x}\rightarrow \mathbf{y}$ is an invertible function between the elements of two sets, with a one-to-one correspondence between elements, ie., each element of one set is paired with precisely one element of the other set and there are no unpaired elements. In the context of probability theory, bijectors are convenient for defining mappings from one probability distribution to another. If a random variate $\mathbf{x}$ with probability density $p(\mathbf{x})$ is transformed under a bijector $\mathbf{f}: \mathbf{x}\rightarrow \mathbf{y}=\mathbf{f}(\mathbf{x})$, then samples and densities transform as:
\begin{align}
    &\mathbf{y} = \mathbf{f}(\mathbf{x}),\; \mathbf{x} = \mathbf{f}^{-1}(\mathbf{y})\nonumber \\
    &p(\mathbf{y}) = |\mathbf{J}|p(\mathbf{x}),\;p(\mathbf{x}) = |\mathbf{J}^{-1}|p(\mathbf{y})
\end{align}
where $J_{ij} = \partial f_i / \partial x_j $ is the Jacobian corresponding to the bijector function $\mathbf{f}$. Any probability density can be represented as a bijection from a base density, or equivalently, any target density can be completely specified by a base density and bijector from that base density to the target. Note also that any composition of bijectors is also a bijector.

In the context of nested sampling, priors $\pi(\btheta)$ over model parameters $\btheta$ should be provided as a bijector that transforms from a base (unit) uniform density $\mathbf{u}\sim\mathcal{U}(0,1)$ to the desired prior $\pi(\boldsymbol\theta)$, ie.,
\begin{align}
    \btheta = \mathbf{f}(\mathbf{u}),
\end{align}
where
\begin{align}
    &\pi(\btheta) = |\mathbf{J}|\,\mathcal{U}(\mathbf{f}^{-1}(\btheta))
\end{align}
In practice, all that is required for specifying nested sampling priors is the bijector function $\mathbf{f}(\mathbf{u})$.

In many use cases, such a bijector is readily available. For example, for any univariate prior density, the bijector from the unit uniform is simply the inverse cumulative density function (CDF) of the target prior,
\begin{align}
    &f(u) = \mathrm{CDF}^{-1}(u), \nonumber \\
    &\mathrm{CDF}(u) = \int_{-\infty}^u \pi(\theta)d\theta,
\end{align}
where the CDF can be interpolated and inverted numerically if the integral is intractable.

However, for the general case of correlated multivariate priors, the requisite bijector is far more involved to compute analytically \citep[see 5.1 of][]{Feroz_2009} and in general numerically requires obtaining by other means. One particularly common scenario where this arises is when one wants to use the (sampled) posterior from one experiment as the prior for another; in these cases, all that is available are samples from the desired prior and un-normalized values of its density at those points.

The solution is to fit a parametric model for the required bijector $\mathbf{f}(\mathbf{u};\mathbf{w})$ (with parameters $\mathbf{w}$) to the target prior. In the following sections we describe how to define expressive parametric bijector models, and fit them to target (prior) densities.
\subsection{Parametric bijectors}
\label{sec:bijector_models}
Parametric bijector models $\mathbf{f}(\mathbf{u};\mathbf{w})$ have recently been gaining popularity for solving complex density estimation tasks, probabilistic learning and variational inference, with a plethora of models available. In this section we give a brief overview of the key methods (in order of increasing complexity). For a more in-depth review, see \citet{papamakarios2019}. 
\subsubsection*{Compositions of simple invertible functions}
Since any composition of bijectors is also a bijector, it is possible to construct flexible parametric bijectors by composing even simple invertible fucntions, such as affine transformations, exponentials, sigmoids, hyperbolic functions, etc. 

Some such transformations applied to a base density lead to already well-studied distribution families, such as the ``sinh-arcsinh" distributions \citep{jones2009sinh} which are generated by applying a $a\,\mathrm{sinh}(b\,\mathrm{sinh}^{-1}(x) + c)$ bijector (with some parameters $a$, $b$ and $c$) to a base normal random variate. Chaining a number of such transformations together, interleaved with linear transformations of the parameter space, can already lead to rather expressive families of distributions.
\subsubsection*{Autoregressive flows}
More sophisticated bijector models typically involve constructing normalizing flows parameterized by neural networks. Of these neural network based models, Inverse Autoregressive Flows (IAFs; \citealp{kingma2016}) have been demonstrated to be effective at representing complex densities, are fast to train, and fast to evaluate once trained. IAFs define a series of $n$ bijectors composed into a normalizing flow, as follows:
\begin{align}
    &\mathbf{z}^{(0)} = \mathbf{n} \sim \mathcal{N}(0, 1) \nonumber \\
    &\mathbf{z}^{(1)} = \boldsymbol{\mu}^{(1)}[\mathbf{z}^{(0)}] + \boldsymbol\sigma^{(1)}[\mathbf{z}^{(0)}]\odot\mathbf{z}^{(0)} \nonumber \\
    &\vdots\nonumber \\
    &\mathbf{y} = \mathbf{z}^{(n)} = \boldsymbol{\mu}^{(n)}[\mathbf{z}^{(n-1)}] + \boldsymbol\sigma^{(n)}[\mathbf{z}^{(n-1)}]\odot\mathbf{z}^{(n-1)}
\end{align}
where the shifts $\boldsymbol{\mu}$ and scales $\boldsymbol\sigma$ of the subsequent affine transforms are autoregressive and are parameterized by neural networks, ie., 
\begin{align}
    \boldsymbol\mu^{(k)}_i = \boldsymbol\mu^{(k)}_i[\mathbf{z}_{1:i-1}; \mathbf{w}],     \boldsymbol\sigma^{(k)}_i = \boldsymbol\sigma^{(k)}_i[\mathbf{z}_{1:i-1}; \mathbf{w}],
\end{align}
where $\mathbf{w}$ denote the weights and biases of the neural network(s). For more details see \cite{papamakarios2017}.
\subsubsection*{Continuous normalizing flows}
In order to take a further step forward in expressivity, one can replace the discrete set of transforms with an integral of continuous-time dynamics \citep{chen2018,grathwohl2018}, i.e., defining the bijector as the solution to an initial value problem:
\begin{align}
    &\mathbf{z}(0) \sim \mathcal{N}(0,1) \nonumber \\
    &\mathbf{y} = \mathbf{z}(t) = \int_0^t \mathbf{f}(\mathbf{z}, t^\prime; \mathbf{w})dt^\prime,
\end{align}
where the dynamics $\mathbf{f}(\mathbf{z}, t; \mathbf{w}) = d\mathbf{z}/dt$ are parameterized by a neural network with weights $\mathbf{w}$. The elevation from discrete to continuous normalizing flows comes with a significant increase in expressivity, making such models appropriate for the most complex bijector representation tasks. See \citet{chen2018} and \citet{grathwohl2018} for more details.
\subsection{Fitting bijectors to target densities}
Fitting a bijector model to the target can be achieved by minimizing the Kullback-Leibler (KL) divergence between the target $\pi(\btheta)$ and the (model) bijective density (with respect to the model parameters $\mathbf{w}$). Since the KL divergence will not be analytically tractable in general, one can instead minimize a Monte Carlo estimate of the integral given samples from the target prior $\{\btheta\} \sim \pi(\btheta)$:
\begin{align}
\label{kl_est}
    \mathcal{L}(\mathbf{w}) = \hat{\mathrm{KL}}(\mathbf{w}) = \frac{1}{N}\sum_{i=1}^N\ln\,\pi^*(\btheta_i; \mathbf{w}),
\end{align}
where $\pi^*(\btheta;\mathbf{w}) = |\mathbf{J}(\btheta;\mathbf{w})|\,\mathcal{U}(\mathbf{f}^{-1}(\btheta;\mathbf{w})$ is the probability density corresponding to the parametric bijector model, with parameters $\mathbf{w}$.

Alternatively, when one has access to not only samples from the target prior, but also the values of the density at those samples, it can be advantageous to regress the model to the target density using the L2 norm, ie., minimize the loss function 
\begin{align}
    \mathcal{L}(\mathbf{w}) = \sum_{i=1}^N ||\,\mathrm{ln}\,\pi(\btheta_i) - \mathrm{ln}\,\pi^*(\btheta_i; \mathbf{w})\,||^2_2.
\end{align}
This has been shown to be less noisy than minimizing the (sampled) KL-divergence in certain cases \citep{seljak2019} (and can be readily extended to exploit gradient information about the target density if it is available; see \citealp{seljak2019} for details).
\section{Cosmological posteriors as bijectors}
\label{sec:cosmological_bijectors}
\begin{figure}
    \centerline{\includegraphics{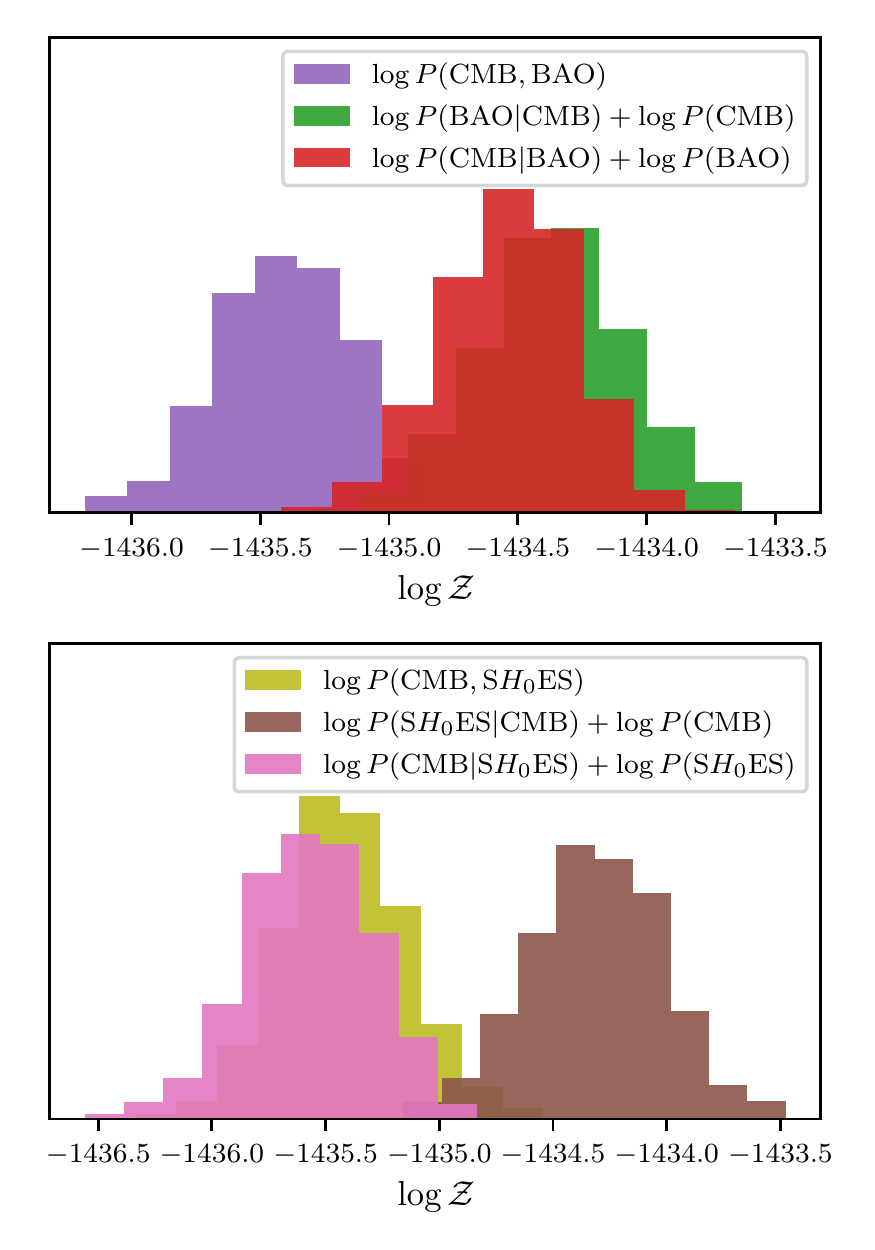}}
    \caption{In the same three paths of Figure~\protect\ref{fig:cosmology_posterior} the Bayesian evidences $\log \mathcal{Z}$ for each of the three runs combine approximately consistently to within error, demonstrating that the bijective prior methodology can be used to recover reasonably accurate evidences in addition to posteriors. It should be noted that in contrast with Figure~\protect\ref{fig:cosmology_posterior}, since the error bars on this plot represent the nested sampling errors in inferring the evidence (in comparison with the posterior probability density spread), one should only expect these plots to overlap to within the histogram width, in analogue with repeated independent measurements of a fixed quantity. We note that typically error estimates on the log-evidence are typically underestimated due to implementation-specific errors~\protect\cite{2019MNRAS.483.2044H} Nonetheless, the inferred evidences are statistically consistent to within $3\sigma$, indicating that there may be a small amount of unquantified error from the use of bijectors, but that the evidences are sufficiently accurate for model comparison purposes (where $\Delta\log\mathcal{Z}\sim\mathcal{O}(1)$). Plot produced under \texttt{anesthetic} \citep{Handley_2019}. \label{fig:cosmology_evidence}
    }
\end{figure}

As a concrete and non-trivial cosmological example we choose the concordance $\Lambda$CDM model of the universe~\citep{scott2018standard} with an additional spatial curvature parameter $\Omega_K$. Such models have been debated recently in the literature \citep{handley2019curvature,Di_Valentino_2019,Efstathiou_2020}. We choose this model not for its controversy, but merely for the fact that it yields non-trivial banana-like priors and posteriors and therefore proves a more challenging cosmological example for a bijector to learn than models without curvature.

For likelihoods we choose three datasets: (1) The \textit{Planck} baseline TT,TE,EE+lowE likelihood (without lensing)~\cite{2020A&A...641A...5P,2020A&A...641A...6P} (hereafter CMB), (2) Baryon Acoustic Oscillation and Redshift Space Distortion measurements from the Baryon Oscillation Spectroscopic Survey (BOSS) DR12~\cite{SDSS,SDSS2,SDSS3} (hereafter BAO), and (3) local cosmic distance ladder measurements of the expansion rate, using type Ia SNe calibrated by variable Cepheid stars from the S$H_0$ES collaboration  ~\cite{Riess2018} (hereafter S$H_0$ES).

The bijector code is implemented as an extension to \texttt{CosmoChord} using \texttt{forpy} to call \texttt{tensorflow} bijectors from FORTRAN.

Figure \ref{fig:cosmology_posterior} shows our key results for CMB and BAO data. In this case we compare the three routes to a combined CMB and BAO constraint on a $k\Lambda$CDM universe. The simplest approach is to run nested sampling with the default \texttt{CosmoMC} prior with both likelihoods turned on $\pi\to$CMB+BAO. The bijector approach first runs the CMB dataset to update the prior $\pi\to$CMB, then after using the posterior samples from this run to train a CMB bijector constructs a distribution to use as the prior for the input for the Bayesian update with the next dataset CMB$\to$BAO. In Figure \ref{fig:cosmology_posterior} we show that doing this in either order recovers the same posterior.

One critical advantage of this approach is that since much of the cost of a nested sampling run is in compressing from prior to posterior, whilst the first run $\pi\to$CMB is expensive, the second update CMB$\to$BAO requires considerably fewer likelihood calls. Furthermore, since the journey from prior to posterior is shorter, less poisson error accumulates over nested sampling iterations, making for more accurate posteriors and evidence estimates. For expensive models like those involving cosmic curvature, this is a significant advantage. 

It should be noted that as CMB and BAO are in mild tension, it has been argued that these combined constraints should be viewed with some scepticism \citep{handley2019curvature,2020arXiv201002230V}, and this tension is much stronger between CMB and S$H_0$ES data~\citep{HubbleTension}. Whilst this is a general cause for simultaneous concern and excitement scientifically speaking, in the context of this work this also serves to highlight the expressivity of these bijector models when combined with nested sampling chains. The tail behaviour of these bijective transformations is sufficiently powerful to still recover the correct answer even when combining likelihoods which are in strong tension. Figure \ref{fig:cosmology_evidence} demonstrates that the evidence estimates extracted by this run from \texttt{anesthetic} \citep{Handley_2019} are also consistent to within the errors on nested sampling.
\section{Conclusions}
\label{sec:conclusions}
Nested sampling implementations have been burdened by a practical limitation that priors need to be specified as bijective transforms from the unit hyper-cube, and in many cases such a representation is not readily available. We have shown that nested sampling priors can be constructed by fitting flexible parametric bijectors to samples from the target prior, providing a general-purpose approach for constructing generic priors for nested sampling. This is of particular importance for use cases where an interim posterior from one experiment is to be used as the prior for another, where a parametric bijector can then simply be trained on samples from the interim posterior. We have demonstrated the use of parametric bijectors on some typical use-cases from cosmology.

In the longer term, we plan to release a library of cosmological bijectors for use as nested sampling priors in order to save users computational resources and standardise cosmological prior choice.
\section*{Acknowledgements}
We thank Johannes Buchner for useful comments. Justin Alsing was supported by research project grant \emph{Fundamental Physics from Cosmological Surveys} funded by the Swedish Research Council (VR) under Dnr 2017-04212.
Will Handley was supported by a Gonville \& Caius Research Fellowship, STFC grant number ST/T001054/1 and a Royal Society University Research Fellowship.
\section*{Data availability}
All the nested sampling runs, code and details for reproducing the results in this letter can be found on Zenodo \citep{Zenodo}.




\bibliographystyle{mnras}
\bibliography{nested_bijectors}



\appendix


\bsp	
\label{lastpage}
\end{document}